# Effect of confinement on flow around a rotating elliptic cylinder in laminar flow regime

Prateek Gupta[1,‡], Sibasish Panda[1,‡], Akhilesh Kumar Sahu[1,*], Deepak Kumar[1]
[1] Department of Chemical Engineering, National Institute of Technology Rourkela-769008, India

[‡]These authors have contributed equally to this work.

*Correspondence: sahuak@nitrkl.ac.in

**Abstract**

The flow phenomena around a rotating elliptic cylinder placed in a narrow channel is studied numerically. The walls of the channel act as a confinement that limits the flow in the transverse direction. The confinement ratio ($\beta$), non-dimensional rotation rate ($\alpha$), and the Reynolds number (*Re*) span across multiple values. A parametric study is done to identify the variations in drag-coefficient ($C_D$), lift-coefficient ($C_L$), and moment coefficient ($C_M$) with changes in $\beta$, $\alpha$ and *Re*. Near-field and far-field vorticity contours are discussed in detail. Fast-Fourier Transform (FFT) of the time-periodic lift signals are presented to understand the shedding-frequency characteristics. Furthermore, $C_M$ values are analysed for possible cases of autorotation. It is observed that confinement acts to delay the shedding of vortices. However, a complete suppression is not obtained even for the maximum value of $\beta$. This is likely because of the sharp flow separation at the edges of the cylinder, which tends to promote the formation of a vortex. Hovering vortices are observed for $\alpha > 1$, and a special case is identified for which the hovering vortex never dissipates.

## Nomenclature

| | |
|---|---|
| $Re$ | Reynolds Number |
| $C_D$ | Drag Coefficient |
| $C_L$ | Lift Coefficient |
| $C_M$ | Moment Coefficient |
| $e$ | Aspect Ratio of the cylinder ($= b/a$) |
| $a$ | semi-major axis, meter |
| $b$ | semi-minor axis, meter |
| $D$ | cylinder diameter, meter |
| $H$ | Vertical height of the channel, meter |
| $L_u$ | Upstream length of the domain, meter |
| $L_d$ | Downstream length of the domain, meter |
| $u_x$ | x-direction velocity, m/sec |
| $u_y$ | y-direction velocity, m/sec |
| $u_\infty$ | freestream velocity, m/sec |
| $p$ | static pressure, Pa |
| $F_D$ | Drag force, N |
| $F_L$ | Lift force, N |
| T | Torque, N-meter |
| $\Delta t$ | time-step size, sec |
| $f^*$ | Non-dimensionalized frequency |
| $T^*$ | Non-dimensionalized time |

*Greek Symbols*

| | |
|---|---|
| $\alpha$ | non-dimensional rotation rate |
| $\beta$ | confinement ratio ($= D/H$) |
| $\mu$ | Dynamic viscosity, Pa-sec |
| $\rho$ | density, Kg/$m^3$ |
| $\omega$ | angular velocity, rad/sec |
| $\tau$ | stress tensor, N/$m^2$ |
| $\theta$ | instantaneous angle of attack, radians |

## Acronyms

| | |
|---|---|
| FFT | Fast-fourier transform |
| CW | Clockwise |
| CCW | Counter-clockwise |
| FVM | Finite volume method |
| SMM | Sliding mesh method |
| LUD | Linear upwind differencing |
| PIV | Particle image velocimetry |
| AOA | Angle of attack |
| AE | Advancing edge |
| AEV | Advancing edge vortex |
| RE | Retreating edge |
| REV | Retreating edge vortex |
| HV | Hovering vortex |
| SF | Shedding frequency |
| RF | Rotating frequency |

## I. INTRODUCTION

The problem of flow past a rotating cylinder of circular and noncircular shape has dictated much of the recent developments in the field of bluff body flows. Having practical applications and rich physics to explore, the current problem can be split into two parts.

The first part can be considered as an elliptic cylinder rotating in an unconfined domain. Immense literature is available on this topic, focusing on both practical applications and flow physics. Iversen [1] gave a description of a Magnus type rotor – the Flettner rotor, which uses the Magnus effect for its rotation. However, the rotation is not self-sustained and requires external torque. This sparked the development of self-sustainable rotors which could auto-rotate, such as the Savonius and Darrieus rotors, often found in wind turbines. Several researchers [2,3] define 'autorotation' as the ability of a device to rotate on its own in the presence of a flow field. However, Lugt [2] argued that this is an incomplete treatment of a complex phenomenon and presented a more comprehensive description for the same. A classic application of 'autorotation' is found in helicopter rotors [3], as the problem of auto-rotating wings helps to understand the concept of 'dynamic stall', a non-linear aerodynamic effect. Many researchers [4,5,6] have documented the use of autorotation mechanics in nature such as winged seeds of certain plants auto-rotating while they fall, in order to increase their dispersal area. Lugt and Ohring [7] used numerical techniques to study the flow physics around a thin rotating cylinder at rest and in a parallel free stream. Lua *et al* [8] used Digital Particle Image Velocimetry (DPIV) to study the flow near a rotating elliptic airfoil and focused mainly on the vortex-shedding behavior with change in Reynolds number and Rossby number ($Ro = 1/\alpha$). Naik *et al* [9] studied the effect of cylinder thickness on the vortex dynamics. Lu *et al* [10, 11] extended the work done by Naik *et al* to also include the effect of $\alpha$ on the vortex shedding patterns and transient force signals at a fixed $Re$. They published two studies covering a wide range of $\alpha$ from 0.5 to 5. More recently, Hu and Tang [12] investigated the effect of $Re$ and $\alpha$ on the far-field wake patterns, and discussed about the autorotative region in their work.

The second part of the current problem can be seen as flow past a rotating circular cylinder in a channel. Presence of a plane surface modifies the flow as it introduces an additional source of shear which can generate vorticity and interfere with the cylinder induced vortices [13]. Another familiar phenomenon associated with these studies is cavitation [14, 15]. The fluid flow near cylinder wall is accelerated due to cylinder's rotation which reduces the local pressure at that point. If this value of local pressure falls below the vapor pressure of the working fluid, cavitation can occur. Gaudemer *et al* [15] used a Couette-based mechanical setup to experimentally study cavitation, whereas Champmartin *et al* [16] has proffered the existence of cavitation based on their numerical findings. Additionally, many researchers study the torque characteristics of the rotating cylinder [16,17,18] to understand the modifications in the flow dynamics due to the presence of walls. An important aspect of a rotating cylinder subject to confinement is its ability to control the flow [19]. This involves suppressing or enhancing vortex-shedding, enhancement in mixing and controlling the aerodynamic forces. Overall, these aspects have been studied under the arc of "hydrodynamic stability" using bifurcation diagrams [20,21].

In this work, we take a step forward from the existing research to understand the flow phenomena near a rotating elliptic cylinder subject to wall confinement. A possible engineering application of this rotating elliptic cylinder is that it can be assumed to be placed upstream of a NACA airfoil in a wind tunnel and can be used to generate turbulent wake structures to evaluate the performance of the airfoil under practical conditions. The best part about a setup of this kind is that the cylinder can be made to autorotate, which means that no external source of torque (i.e., a motor) will be needed which will save some electricity. However, caution should be taken since this study reports results only for incompressible flow, which cannot be translated directly into an experiment with a compressible working fluid. To the best of the authors' knowledge, very limited literature is available on flow around a stationary elliptic cylinder in confinement, and none for a rotating one.

## II. LITERATURE REVIEW

The study of fluid-flow around a circular cylinder has been one of the most researched topics in the sphere of bluff body flows with studies dating back to the 19th century. However, the study of fluid flow around an elliptic cylinder is scant in comparison to its circular counterpart. The evolution timeline of research starting with the flow around an unconfined circular cylinder to study of flow around elliptic cylinders under the effect of rotation, and their findings have been presented in the brief literature review.

One of the first comprehensive review of flow past a steady circular cylinder has been reported by Zdravkovich [22,23], which highlighted vital flow kinematics such as the range of Reynolds number for flow separation, the increase of wake Length, and transition of flow with Reynolds number. Williamson [24] discussed the contemporary developments for the study and analysis of vortex dynamics in cylinder wake in a detailed review. With the advancement in technologies and their subsequent adaptation into the research fraternity, the research being undertaken have become more intricate accounting for multiple relevant research parameters such as vortex shedding frequency, asymmetry, and blockage ratio in confined cases, eccentricity for elliptic cylinders, and rotation rate for flow dynamics further downstream. The fluid flow in a channel over a cylinder, such as the current study, can be categorized as both internal and external flow.

In a low Reynolds number study for flow past circular cylinder placed between parallel walls, Singha and Sinhamahapatra [25] reported a positive dependence of critical Reynolds number, drag coefficient, Strouhal number, and an inverse dependence of lift coefficient on blockage ratio. Moreover, they also reported that Strouhal number does not depend on Reynolds number at low blockage ratios. Cliffe and Tavener [20] used Direct Numerical Simulations to study the effect of blockage and confinement on periodic flows, in which they described the flow physics based on the presence of Hopf bifurcation points and the plane of symmetry. They reported restabilization of steady flows at large blockage ratios with an increase in Reynolds number even for non-rotating cylinders. Later, a similar study (DNS) of confined flow physics in low Reynolds number regime by Ooi et al [26] elaborated the behavior of fluctuating lift forces over a range of blockage ratio, increase in critical transition Reynolds number with blockage ratio between steady and unsteady flow as well as two and three-dimensional flows. The effect of channel confinement and shear of the incoming flow was studied in detail by Zovatto and Pedrizzetti[13]. The results were analyzed both qualitatively and quantitatively to determine the geometric and parametric influence on flow physics and force coefficients. Champmartin et al [16] studied the effect of introduction of cylinder rotation in confinement and reported the dependence of torque on blockage ratio as well as its asymmetry in presence of a single plane, the existence of backflow for both planes as well as the validity of reciprocity theorem in presence of boundaries. A study by Paramane et al [19] to determine the effect of channel-confinement and rotation on a laminar flow revealed a novel crisscross movement of shed-vortices in the flow downstream. Flow transition in the study was reasoned on entrainment of the fluid below the cylinder into the wake region and the interaction of vortices generated near the cylinder. An effort to obtain critical Reynolds number and non-dimensionalized rotational velocity was made to specify the steady-unsteady flow transition. The study effectively concludes on the stabilizing effect of rotation and channel-confinement based on results obtained on behavior of shear-layer, wake region, Strouhal number, lift and drag coefficients for various blockage ratios and rotation rates. In a study of power-law fluid flow round a rotating and confined cylinder, Thakur et al [18] reported an increasing trend in torque with increasing Reynolds number, blockage, and symmetry, and a higher value of torque was observed for confined and Newtonian than unconfined and shear-thinning fluids respectively.

Introduction of further complexity into the study of flow dynamics past an unconfined steady and rotating elliptic cylinder demonstrates flow transition as well as rich wake physics. An analytical approach adopted by Khan et al [27] for the study of fluid flow and heat transfer from elliptic cylinders used the Von Karman–Pohlhausen integral method for both fluid and thermal boundary layer analysis. The formulated correlations showed that the drag coefficients are low and the heat transfer rate are higher for elliptic cylinders than their circular counterparts. The dependence of drag coefficient and average heat transfer rate on Reynolds number and aspect ratio was also presented. Sen [28] described the flow phenomena in terms of surface pressure and viscous forces for a steady elliptic cylinder inclined at different angles of attack. The study made an effort to determine the stagnation points on the cylinder surface in order to elaborate the flow separation. The nature of the variation of hydrodynamic parameters such as drag and lift coefficients was studied with respect to Aspect ratio and Reynolds number. In a numerical study of Newtonian and power-law fluid flow around an elliptical cylinder, Sahu et al [29] elaborated positive dependence of the Critical Reynolds number on eccentricity and inverse dependence on the power-law index. Rich wake physics with flow transition, which becomes obvious as the aspect ratio decreases with an increase in Reynolds number, is elucidated in a review paper by Thompson et al [30]. Furthermore, the paper highlights complications in centerline wake deficit, Strouhal number variation, and deviation from Bénard–von Kármán vortices in comparison to a circular cylinder. Zhu and Holmedal [31] discussed flow separation around a partial confined elliptic cylinder with the help of oscillating stagnation and separation points over both steady and unsteady flow regimes.

An elaborate experimental study using DPIV measurements was carried out by Lua *et al* [8] for two flow configurations-rotation of an elliptic cylinder in fluid at rest and parallel freestream. For rotation in fluid at rest the vortex suction effect dampens the torque and two distinct vortex formations were observed with Reynolds number. For rotation in freestream, a "Hovering vortex" was reported for the very first time and the rich vortex dynamics was studied with the change in cylinder tip-freestream velocity ratio. A numerical study of flow past a rotating elliptic cylinder by Naik *et al* [11] focused on the orientation, strength as well as the frequency of shed vortices and its dependence on eccentricity and rotation rate using an immersed boundary method. The lift coefficient was observed to increase with an increase in rotation rate and decrease in eccentricity whereas a complex dependence was observed for the drag coefficient. Vortex shedding frequency was highlighted to be a strong function of rotation rate. A numerical study in an unconfined domain by Lua *et al* [9] demonstrated the fundamental dependence of drag and lift coefficients on evolution and positioning of vortices as well as dependence on rotation velocity, eccentricity stress on physical aspects such as thrust generation, its inception, and their probable dependence on Magnus effect. Subsequently, the study [10] has been expanded with a focus on dependence and variation of Aerodynamic forces with flow parameters such as high rotation rates and eccentricity along with insights on the evolution of wakes, rotating, and hovering vortices. Previous works on confined rotating circular cylinders in power-law, as well as an unconfined rotating elliptic cylinder with varying rotating rates and blockages, provide` a deep insight into the flow characteristics and rich wake physics. The present work tries to study the effect of confinement and cylinder rotation on the flow phenomena. The parameters considered are: $e = 0.1$, Reynolds number $50 \leq Re \leq 150$ blockage ratio ($\beta = 1/2, 1/4, 1/6$ and $1/8$) and rotation speed ($\alpha = 0.5, 1$ and $2$).

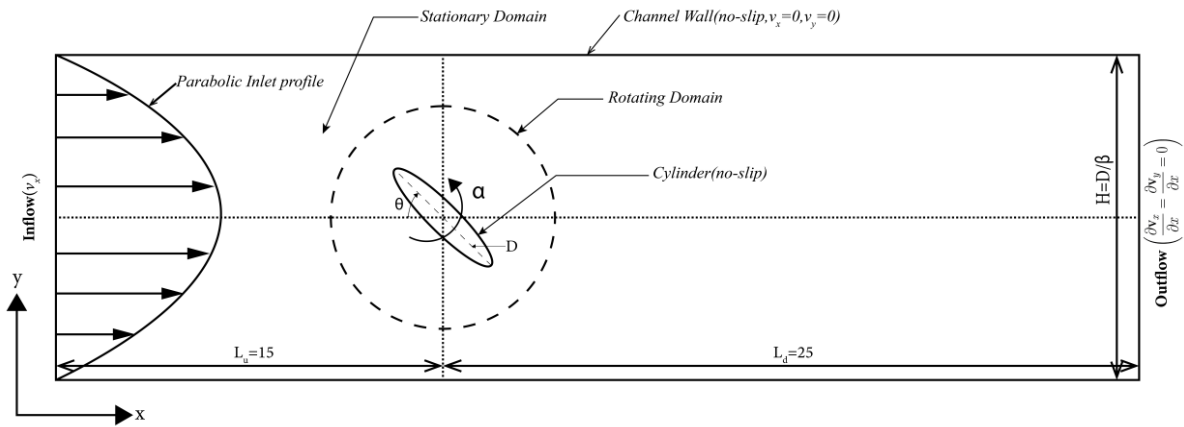

Fig 1: Schematic representation of computational domain.

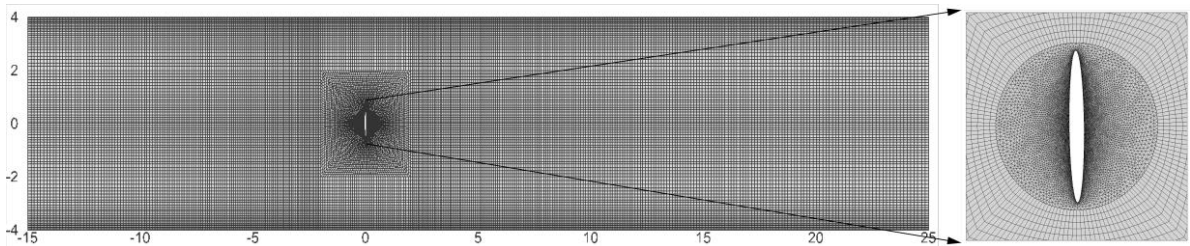

Fig 2: Schematic representation of Multi-block mesh used in the study.

### III. PROBLEM DESCRIPTION AND NUMERICAL METHODOLOGY
#### A. Problem formulation

The present work studies the effect of channel confinement on flow past an elliptic cylinder rotating in a parallel stream, with its axis perpendicular to the direction of the flow (Fig 1). It is assumed that the cylinder is infinitely long in the span-wise direction to avoid any three–dimensional effects. Also, the maximum Reynolds number ($Re$) of the flow and the highest value of non-dimensional rotation rate ($\alpha$) are such that we can safely discount the possibility of any three-dimensional perturbations [26]. The cylinder is characterized by its aspect ratio ($e$), which is defined as the ratio of its semi-minor ($b$) to semi-major ($a$) axis and is considered 0.1 for this study. The cylinder rotates counter-clockwise (CCW), and the angle between its major axis and the direction of freestream is defined as the instantaneous angle of attack (AOA). The Reynolds number of the flow is defined as ($Re = \frac{\rho D \overline{u_x}}{\mu}$), where $\overline{u_x}$ is the average inlet velocity, and $D$ is the cylinder diameter ($= 2a$). The rotational speed of the cylinder is characterized by a non-dimensional rotation rate ($\alpha = D\omega/2U_x$), where $\omega$ is the angular velocity in rad/sec.

The governing continuity and momentum equations can be written as:

Continuity equation

$$\frac{\partial u_x}{\partial x} + \frac{\partial u_y}{\partial y} = 0 \qquad (1)$$

Momentum equation

$$\rho\left(\frac{\partial u_x}{\partial t} + \frac{\partial (u_x u_x)}{\partial x} + \frac{\partial (u_x u_y)}{\partial y}\right) = -\frac{\partial p}{\partial x} + \left(\frac{\partial \tau_{xx}}{\partial x} + \frac{\partial \tau_{yx}}{\partial y}\right) \qquad (2)$$

$$\rho\left(\frac{\partial u_y}{\partial t} + \frac{\partial (u_x u_y)}{\partial x} + \frac{\partial (u_y u_y)}{\partial y}\right) = -\frac{\partial p}{\partial y} + \left(\frac{\partial \tau_{xy}}{\partial x} + \frac{\partial \tau_{yy}}{\partial y}\right) \qquad (3)$$

Where $u_x, u_y, \rho, p, \tau_{ij}$ denote the fluid velocity in $x$ and $y$ velocity, density, pressure and extra stress tensor respectively.

The boundary conditions at the domain boundaries and cylinder wall for present problem are written as follows:

- At the inlet plane a parabolic flow profile in $x$- direction and a zero velocity in y-direction can be written mathematically as

$$u_x = U_\infty^2\left(1 - \frac{y^2}{H^2}\right) \text{ and } u_y = 0 \qquad (4)$$

- On the surface of the cylinder and Upper and lower wall: No-slip boundary condition and normal velocity, is applied. They can be written as

$$\frac{\partial u_x}{\partial y} = 0; \ u_y = 0 \qquad (5)$$

- At the outlet boundary: The outflow boundary condition is used. The outflow condition corresponds to zero diffusion fluxes in the direction normal to the outlet planes. It is used for all the dependent variables.

$$\frac{\partial u_x}{\partial x} = 0; \ \frac{\partial u_y}{\partial x} = 0 \qquad (6)$$

Drag, Lift and Moment coefficients have been used in the study to comprehend and analyze the flow physics quantitatively. They are defined as

$$C_D = \frac{F_D}{(1/2)\rho U_\infty^2(2a)}, \ C_L = \frac{F_L}{(1/2)\rho U_\infty^2(2a)}, \ C_M = \frac{T}{(1/2)\rho U_\infty^2(2a)^2} \text{ and } C_P = \frac{P-P_\infty}{(1/2)\rho U_\infty^2} \qquad (7)$$

The cylinder is confined between two plane walls separated by a vertical distance *H*, with each wall at a distance of *H/2* from the geometric center of the ellipse. Following this, we define the blockage $\beta$ as *D/H*. A User Defined Function (UDF) was compiled and implemented at the inlet boundary to ensure that the velocity profile is parabolic in nature right from the beginning. Given that the walls are finite in length; a parabolic profile ascertains that the flow has fully developed before it reaches the cylinder.

### B. Numerical methodology

We use finite-volume method (FVM) based commercial software ANSYS FLUENT to numerically solve the governing equations. The entire domain is divided into two zones (Fig. 2a), namely the 'rotating domain' and the 'stationary domain'. The rotating domain is circular in shape and is meshed using triangular cells which gives more flexibility and economy over cell distribution, with higher mesh density near the cylinder boundary. The stationary domain is made up of a multi-block structured grid to minimize numerical diffusion (Fig. 2b). To induce cylinder rotation, Sliding Mesh Method (SMM) is implemented, wherein the rotating domain moves relative to the stationary region. Full pressure-velocity coupling is achieved by using a pressure-based coupled solver, which results in faster convergence. The convective term in the flow equation is discretized using Linear Upwind Differencing (LUD) which uses a three-point stencil to give second order accuracy, whereas the approximation for diffusion term is automatically second-order accurate. Discretization with respect to time is first-order implicit which magnifies the stability envelope several-fold. The time-intensive nature of the coupled algorithm is compensated by using an already developed flow field as input for the simulations. This improvisation reduces the total simulation time dramatically, without affecting solution accuracy. A convergence criterion of $10^{-6}$ is set for the residuals of continuity and velocity terms, since it is sufficient to ensure convergence. We ran a test case for 125 time-steps with 35 iterations per time-step. This ensured that the continuity residual dipped to the $10^{-6}$ mark at each time-step, i.e., a converged solution was obtained this way for each time-step. We monitored the drag coefficient and averaged it to get a value of 3.79511. For the second test case, the number of iterations per time-step was increased to 50. This again ensured that the residuals dipped to the required $10^{-8}$ mark at every time-step. Taking the average value over 125 time-steps, we obtained a drag coefficient value of 3.79511 again. Thus, an exactly same value of $C_D$ was obtained in both cases, thereby confirming that a convergence criterion of $10^{-6}$ is sufficient.

In confined studies of flow around bluff bodies the choice of appropriate domain, grid and time-step size plays a crucial role in precision of obtained results. The accuracy of our results has been ensured by performing the independence tests.

### C. Domain independence study

A sufficiently long domain in comparison to its breadth is chosen to capture the flow field. Aerodynamic parameters such as drag and lift coefficients are compared so as to get a quantitative idea of domain size effect. The domain independence study is performed at Reynolds number $Re = 150$, dimensionless rotational speed $\alpha = 0.5$ and blockage $\beta = 1/8$ to ensure robustness of the domain. It can be reasoned that for the case of elliptic cylinders' maximum disturbance is observed at this case due to formation of the highest magnitude of vortices and hence this case is chosen as the test case. As the parabolic flow inlet ensures a fully developed flow profile, the upstream length was fixed at $L_u=15$. The downstream length was varied keeping the upstream length fixed, downstream length $L_d=25$ was found to be optimal for running the simulations.

**Table I: Domain size independence test**

| Upstream length ($L_u$) | Downstream length ($L_d$) | Drag coefficient ($C_D$) | Lift coefficient ($C_L$) |
|---|---|---|---|
| 15D | 20D | 2.256 | -1.855 |
| 15D | 25D | 2.261 | -1.858 |
| 15D | 30D | 2.255 | -1.854 |

### D. Grid independence study

Subsequently, to attain the optimal grid, simulations were run on an estimated grid. The number of nodes were both doubled and halved to attain the best suited grid. Details of the three meshes created for this purpose have been presented in Table 2, which summarizes the influence of mesh characteristics, such as number of nodes, on aerodynamic parameters such as lift and drag. The rotating domain has been meshed using the face sizing function and the interface edge has been divided into equal parts so as to facilitate the smooth transition of information from the stationary to rotating domain. Accounting for the negligible change in aerodynamic parameters and a significant increase in the computational time from G2→G3, G2 has been finalized as the mesh to be used for the study.

**Table II: Grid independence test**

| Grid | No. of Nodes | No. of Elements | Drag coefficient ($C_D$) | Lift coefficient ($C_L$) |
|---|---|---|---|---|
| G1 | 24651 | 27097 | 2.202 | -1.816 |
| G2 | 43577 | 50074 | 2.261 | -1.858 |
| G3 | 91796 | 103932 | 2.275 | -1.869 |

### E. Time independence study

Based on the physical properties of the fluid as well as flow conditions, the flow gradually develops time-dependency and periodicity. Precision of results in such time-dependent studies is heavily dependent on the choice of time-step size. The time independence test was carried out for three different time step size $\Delta t =$ 0.0005, 0.001 and 0.002. Table 3 demonstrates value of aerodynamic parameters obtained by time-integration of field equations with different time-steps. It is seen that the values show a negligible difference as we move from $\Delta t=0.001$ to $\Delta t=0.0005$. Hence, time-step size of $\Delta t = 0.001$ is deemed optimal for use without affecting the accuracy of results.

**Table III: Time-step size independence test**

| Time-step size | Drag coefficient ($C_D$) | Lift coefficient ($C_L$) |
|---|---|---|
| 0.002 | 2.232 | -1.817 |
| 0.001 | 2.261 | -1.858 |
| 0.0005 | 2.274 | -1.879 |

To summarize the section, the results obtained in this work are based on the following domain, spatial and temporal parameters: $L_u = 15$, $L_d = 25$, grid G2 and $\Delta t = 0.001$ for the periodic flow regime.

The variable parameters used for the current numerical investigation is done for the following parameters space:

Reynolds number $Re$ = 50, 100, 150

Rotational speed $\alpha$ =0.5, 1, 2

Blockage ratio $\beta = \frac{1}{2}, \frac{1}{4}, \frac{1}{6}$ and $\frac{1}{8}$

## IV. RESULTS
### A. Validation

To establish the reliability of the chosen numerical methodology, a few validatory studies have been performed. Table 4 lists and compares the present results with the literature values. Lua *et al* [9] used the PIV technique to analyze the flow phenomena around a rotating elliptic cylinder and reported vorticity contours (scaled by rotational frequency). The force coefficients are compared with the values reported by Lua *et al* [9] and the agreement is excellent. To check the confined flow computations, comparisons are made with results from the recent numerical work by Thakur *et al* [18] on confined flow around rotating circular cylinder at low Reynolds number. Also, the validations at comparable Reynolds number and rotation speeds are presented here with confined flow over a rotating circular cylinder by Paramane *et al* [19]. The maximum and minimum deviation in drag coefficient value is 2.67% and 0.3% respectively. On the other hand, the maximum and minimum deviation in lift coefficient value is 3.33% and 0.02% respectively. The observed deviations are well within the acceptable error limit for a numerical study and hence the chosen numerical methodology is deemed suitable for current study.

**Table IV: Comparison of present results with Literature**

| Parameters | | | | $C_D$ | | $C_L$ | |
|---|---|---|---|---|---|---|---|
| $\beta$ | $e$ | $Re$ | $\alpha$ | Literature values | Present values | Literature values | Present values |
| 0.5 | 1 | 100 | 1 | 5.960 [18] | 5.978 | -4.397 | -4.366 |
| 0.5 | 1 | 70 | 2 | 7.66 [18] | 7.6338 | -9.834 | -9.898 |
| 0.2 | 1 | 40 | 1 | 3.2801 [17] | 3.2882 | -3.6409 | -3.7220 |
| 0.6 | 1 | 40 | 1 | 16.287 [17] | 16.384 | -6.770 | -6.788 |
| - | 0.6 | 200 | 1.5 | 0.954 [9] | 0.936 | -2.868 | -2.889 |
| - | 0.125 | 200 | 0.5 | 1.197 [9] | 1.165 | -1.263 | -1.221 |
| - | 0.2 | 200 | 2 | 0.7 [9] | 0.705 | -2.684 | -2.664 |

## B. Streamlines

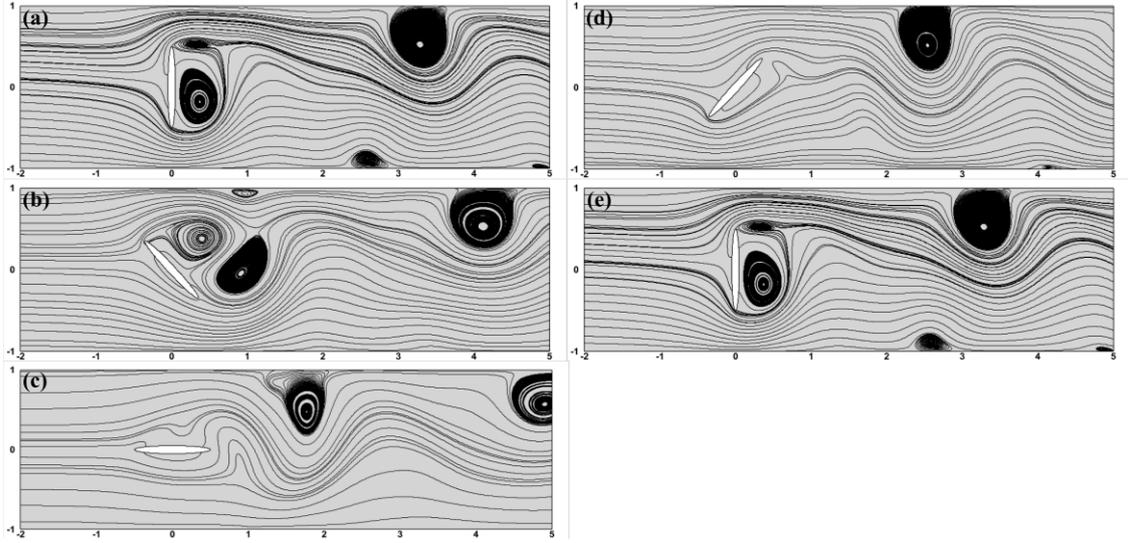

Fig 3: Streamlines of $Re = 150$, $\beta = 1/2$, α = 0.5.

In this section, we present a brief discussion on streamlines to address the instantaneous velocity field distribution. These streamlines are nothing but a family of curves, and at any point on these curves, a velocity vector can be drawn tangentially to represent the velocity field. For $Re = 150$ and $\alpha = 0.5$, Figure 3 and 4 display the instantaneous streamlines for $\beta = 1/8$ and $\beta = 1/2$ respectively. It should be noted that it takes $2\pi$ rotation of the cylinder for the flow to repeat itself at $\beta = 1/8$, and $\pi$ rotation at $\beta = 1/2$.

From Figure 3, we can see that multiple vortices are formed in the downstream wake. In Fig 3(a) and 3(b), we can clearly see a vortex very close to the leeward side of the cylinder, thereby creating a strong suction on that side of the cylinder. This results in a negative $C_P$ value for the leeward side (shown as the dotted surface-1 in Fig. 10). In general, several critical points, viz. saddles and nodes (terminology from Hunt *et al* [33]) are observed in the flow. Moving laterally to the flow direction, the turbulence in the flow vanishes beyond Y = ±1, and the streamlines become nearly straight as we approach the walls. Figure 4 tells a different story; as we move laterally towards the walls, a large vortex is observed on the upper wall. Also, in Figure 4(a) and 4(b), we notice that the leeward surface of the cylinder harbours a negative pressure zone due to presence of vortical structures. This finding can again be validated from the $C_P$ curves (Fig. 10). Additionally, it should be noted that the number of critical points in the downstream flow is lesser in comparison to $\beta = 1/8$.

## C. Vorticity Contours

To understand the vortex shedding phenomena, the flow field is analyzed with the help of vorticity contours. The contours are presented here after attaining the periodic flow. The vorticity values are non-dimensionalised by rotational frequency of the cylinder and range from -30 to +30 in all the figures discussed in this section. The contours are presented for half rotation period at intervals of $\frac{\pi}{4}$ rad $\left(\text{AOA} = \frac{\pi}{4}, \frac{\pi}{2}, \frac{3\pi}{4}, \pi\right)$. In some figures they are presented for one rotation period so as to show the periodicity of the fluid flow around the cylinder.

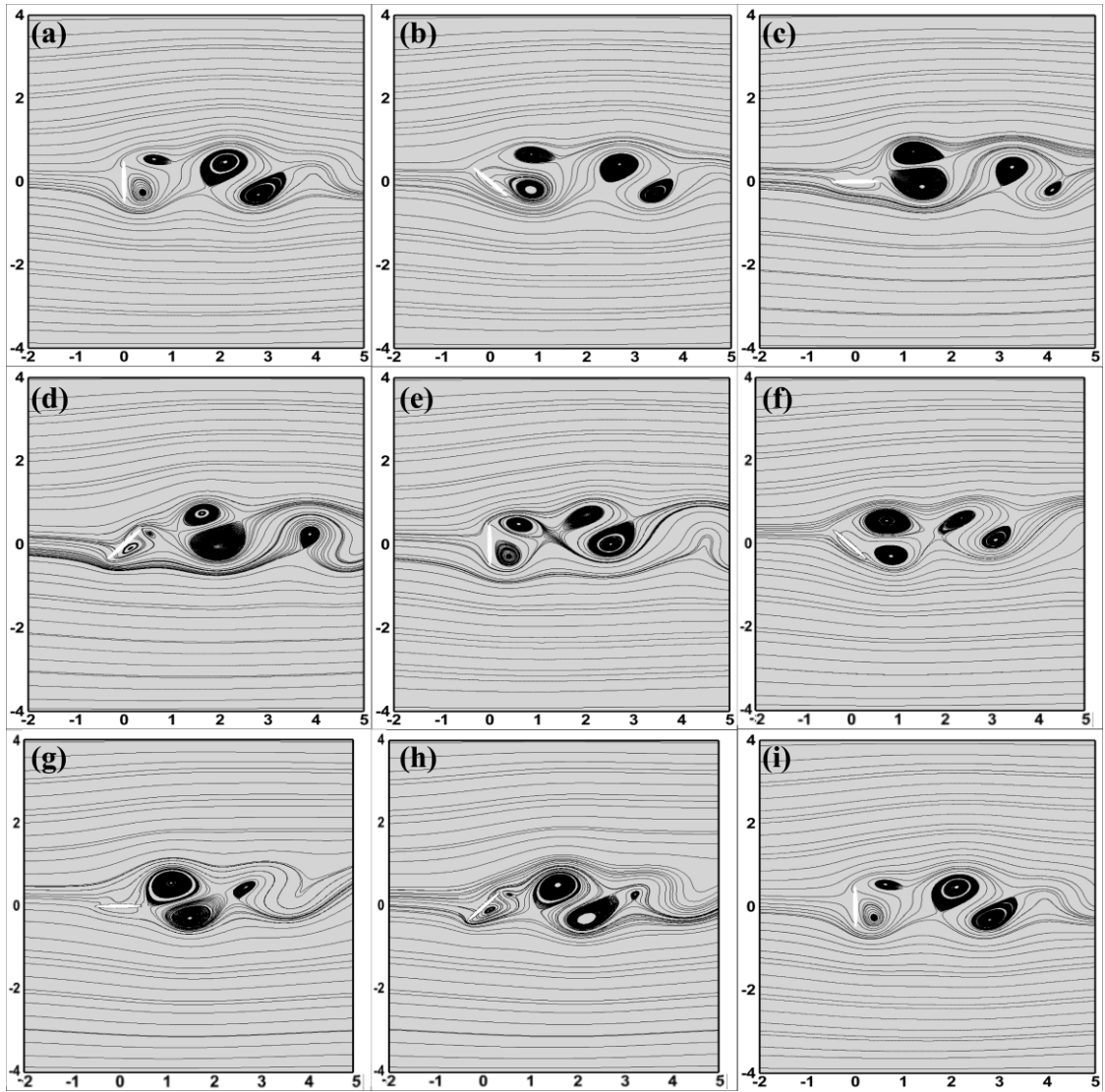

Fig 4: Streamlines for Re = 150, $\beta = 1/8$, $\alpha = 0.5$.

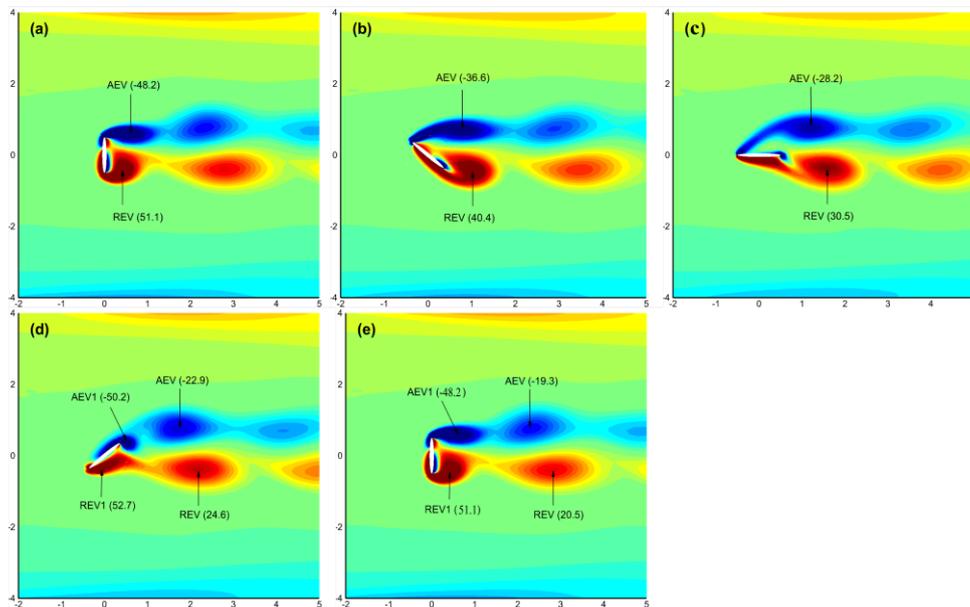

. Fig 5: Vorticity Contours of $Re = 50$, $\beta = 1/8$, $\alpha = 0.5$ for a single vortex shedding period.

Figure 5 shows the vorticity contours for $\beta = 1/8$ at $\alpha = 0.5$ and $Re = 50$. Figure 5a demonstrates the reference orientation of the cylinder to incoming flow. Here, the upper edge is termed as Advancing edge and the lower edge is termed as Retreating edge. The flow separation at the edges leads to the formation of an Advancing edge vortex (AEV) and Retreating edge vortex (REV). It can be seen in Figure 5(b) that the vortices (AEV and REV) are still attached to the cylinder after the interval of $\frac{\pi}{4}$ rad. As the cylinder rotates further, both the AEV and REV are detached (Fig 5c) and subsequently shed (Fig 5d) in the downstream direction. A symmetric reduction in vortex strength (implied by the vorticity values) is observed as the value of both AEV and REV reduces by approximately 60% over a complete vortex shedding cycle. Figure 5e demonstrates repetition of vortex shedding pattern after half rotation period i.e. the cylinder rotates by $\pi$ radians. We can also say that a pair of vortices sheds in every half rotation period and it can be seen that the pair looks almost same as it convects downstream.

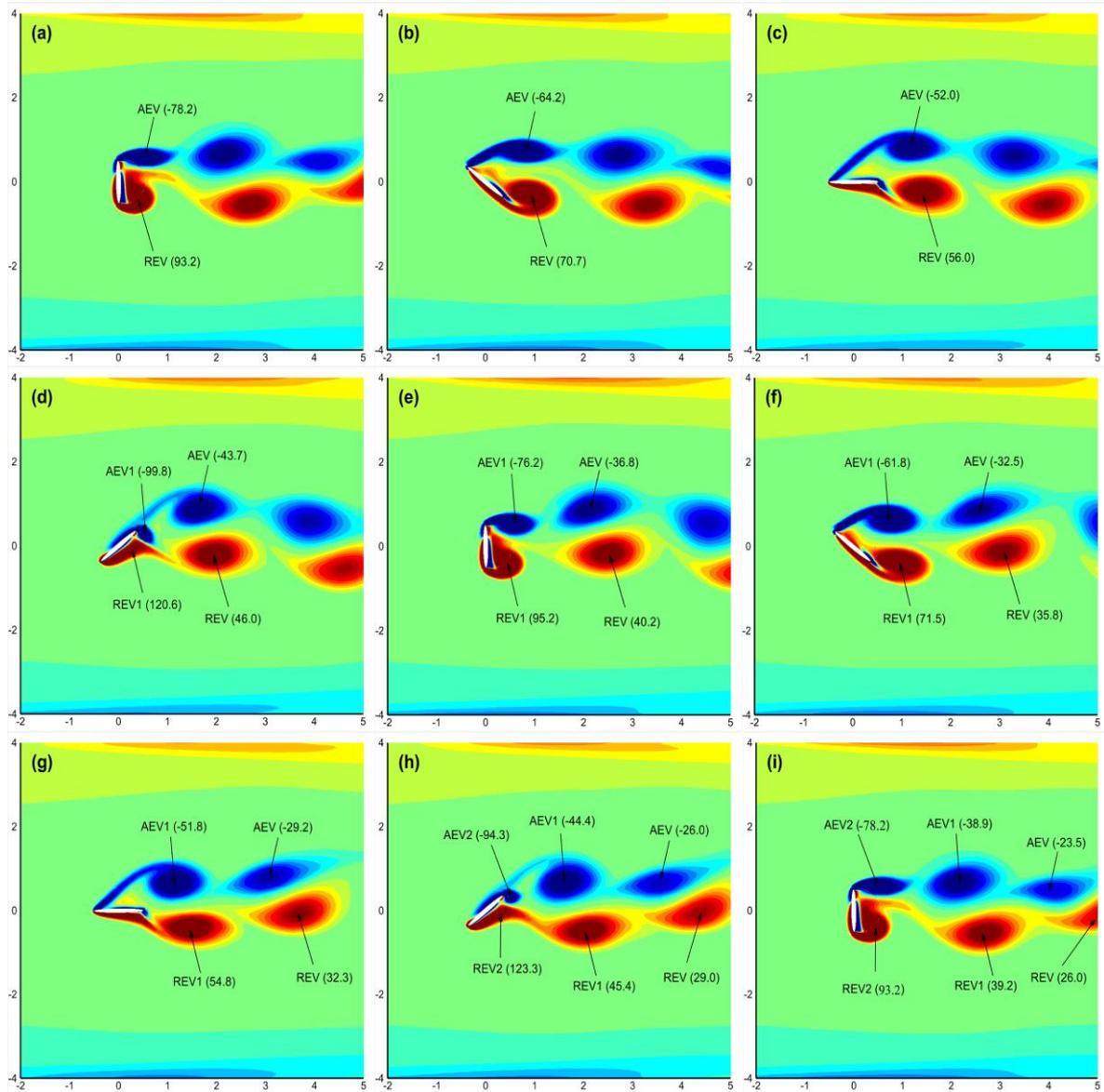

Fig 6: Vorticity Contours of $Re = 150$, $\beta = 1/8$, $\alpha = 0.5$ for a single vortex shedding period.

Figure 6 shows the vorticity contours in the vicinity of cylinder at $Re = 150$. With the help of maximum value of vorticity magnitude of AEV and REV from Fig 5(a) and 6(a), we hypothesize that the increased value in Figure 6(a) is due to increased flow inertia. An interesting feature to note is the change in the periodicity of the vortex shedding. Even though vortices are shed every half rotation, it takes one complete rotation or $2\pi$ radians for the vortex shedding pattern to repeat. Two distinct pair of vortices are shed over each complete rotation. At $Re = 50$, the fluid flow in the vicinity of cylinder is repeated after half rotation period (Fig 5(a) and (e)) whereas it takes one rotation period to reach periodicity at $Re = 150$ (Fig 6(a) and (i)). Also, a slower convection speed can be observed, as the vortices take more time to cross the $x = 5a$ length in the streamwise direction despite the increased flow inertia.

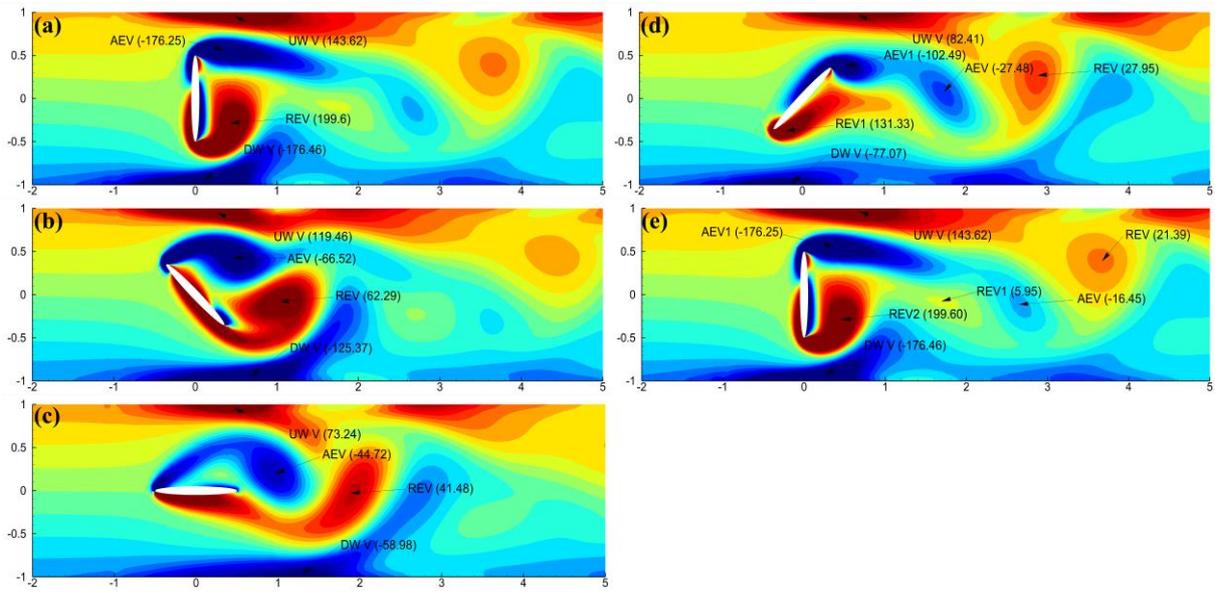

Figure 7: Vorticity Contours of $Re = 50$, $\beta = 1/2$, $\alpha = 0.5$ for a single vortex shedding period.

As the blockage ratio is increased, intense flow separation occurs at both edges (Fig 7). Also, the increased fluid velocity in the gap between cylinder and wall due to increased blockage leads to formation of prominent wall shear layers. The maximum vorticity of AEV and REV is significantly higher in magnitude than same observed for $\beta = 1/8$. It is observed that the cylinder vortex interacts with channel wall vortex. Also, the AEV deflects towards the lower wall and REV go towards upper wall while flowing in downstream direction. The deflection occurs when the distance between cylinder edge (top or bottom) and wall is minimum.

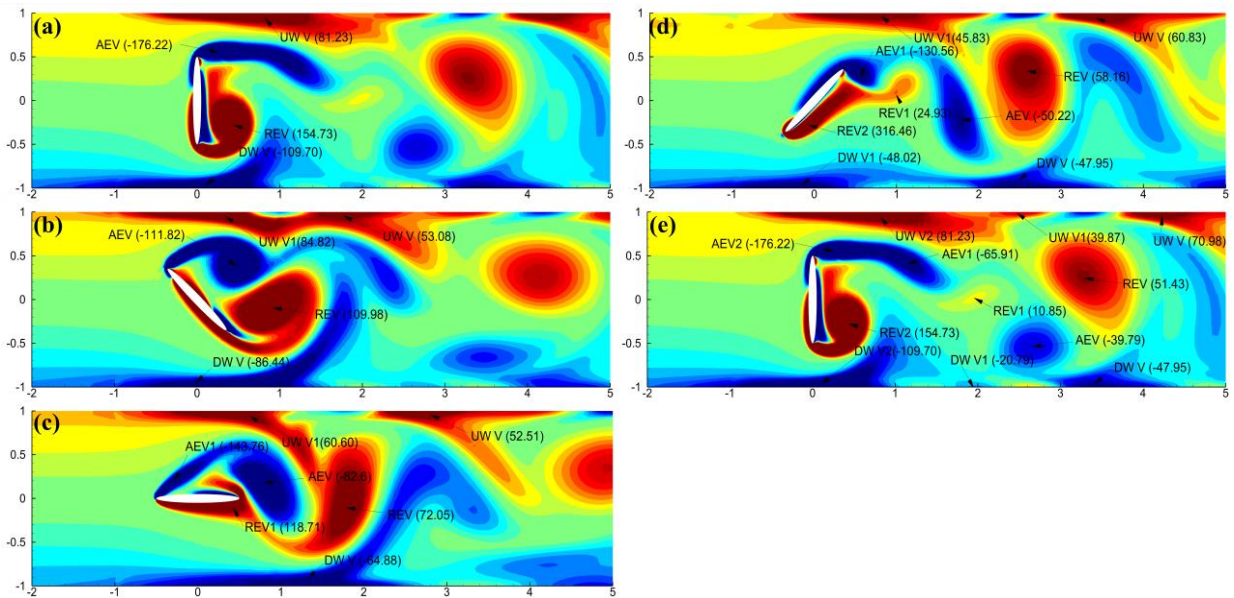

Fig 8: Vorticity Contours of $Re = 150$, $\beta = 1/2$, $\alpha = 0.5$ for a single vortex shedding period.

Instantaneous vorticity contours at $Re = 150$ are shown for blockage $\beta = 1/2$ in Figure 8. Here also, the vorticity patterns repeat every half rotation cycle or $\pi$ radians. The interaction between edge vortex and wall vortex is more pronounced here as there is shedding of vortices from both the walls. Also, the upper and lower wall vortices shed and merge with REV and AEV respectively every half rotation cycle.

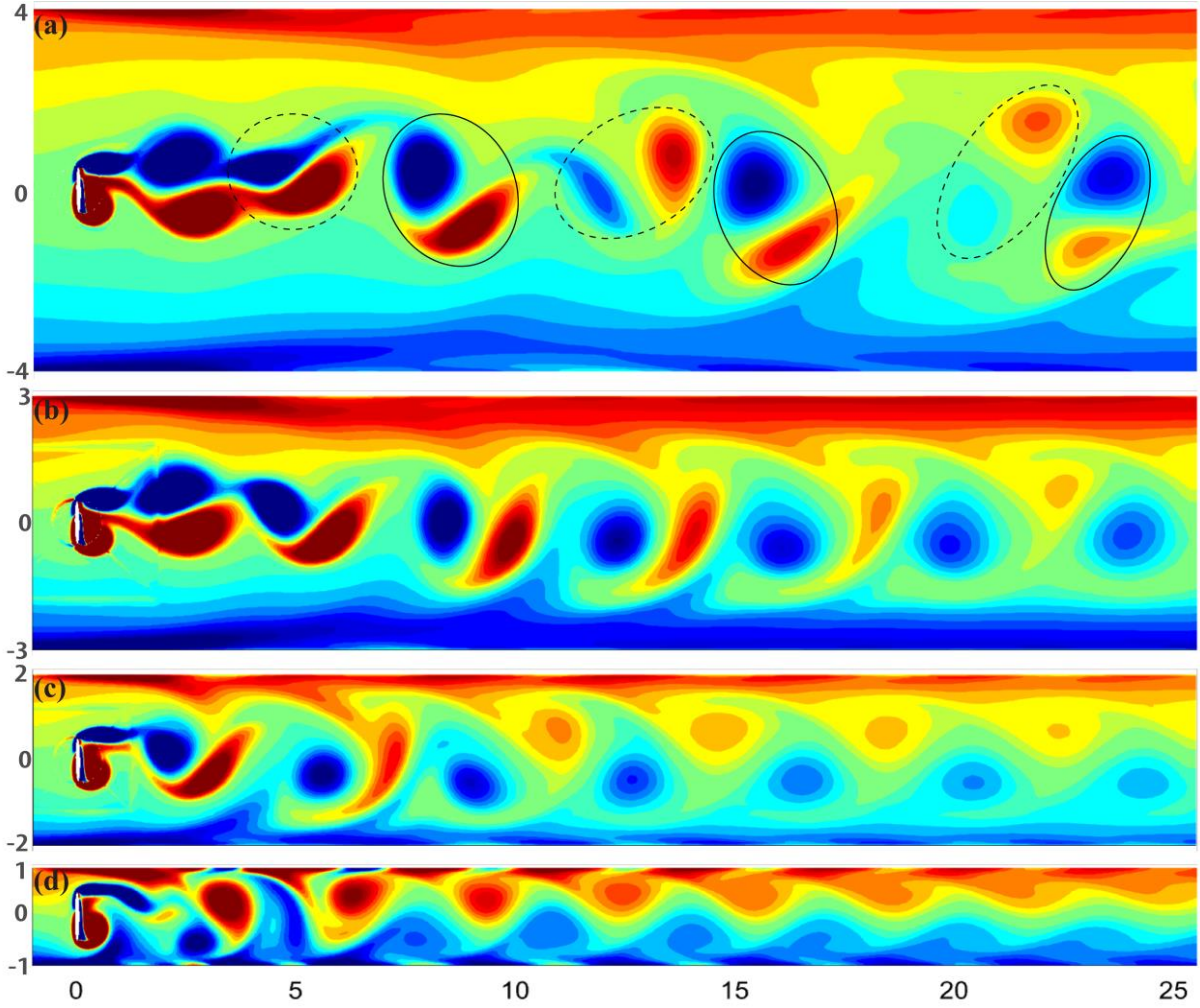

Fig 9 : Far-field vorticity contours for all blockages at $Re = 150$ and $\alpha = 0.5$.

Figure 9 shows the far-field vorticity contours inside the full downstream length of the computational domain for $Re = 150$, and $\alpha = 0.5$ at different values of $\beta$. At this particular Reynolds number and rotational rate value, the likelihood of vortex-shedding for an unconfined domain is highest, which is why we use this case to demonstrate the suppressive effect that blockage has on vortex shedding. At $\beta = 1/8$, the downstream wake consists of set of two pairs of vortices. The vortex pair enclosed by the dashed line is hereby termed Pair-I, and the other pair enclosed by the solid line is Pair-II. Every half-rotation period, one pair is shed from the cylinder. It is seen that the AEV for Pair-I dissipates very rapidly in comparison to its complementary REV. On the other hand, the REV for pair-II shreds faster than its AEV. The far-field pattern for $\beta = 1/6$ shows a row of CW (blue) vortices which appear to be axis-symmetric as they move downstream. And these perfectly round CW (negative) vortices are interspersed with CCW (positive) vortices which in turn are stretched out and interact very strongly with the same sign wall vorticity. We can see that the REVs shed from the cylinder deflect towards the upper wall as the x-coordinate increases. A $\beta$ value of 1/4 appears to accelerate the developments taking place for $\beta = 1/6$. It is observed that the REVs getting deflected towards the upper wall after being stretched out, whereas, the AEVs tend to move towards the lower wall after enjoying a brief period of seclusion. Finally, at $\beta = 1/2$, the far-field is completely devoid of any noticeable vortex eye. This is because the shed vortices have completely merged with the same sign wall vorticity. The distance between the shear layers of the two walls at this value of blockage is so small that it is unable to accommodate a vortex. We thus clearly see how blockage leads to a suppression of vortex shedding in the downstream.

Figure 10 shows variation of area-weighted pressure distribution for a quantitative understanding of the wake. Owing to the maximum blockage of the channel at a cylinder orientation of $\frac{\pi}{2}$ rad, a maximum pressure deficit is observed at $\beta = 1/2$ and minimum deficit at $\beta = 1/8$.

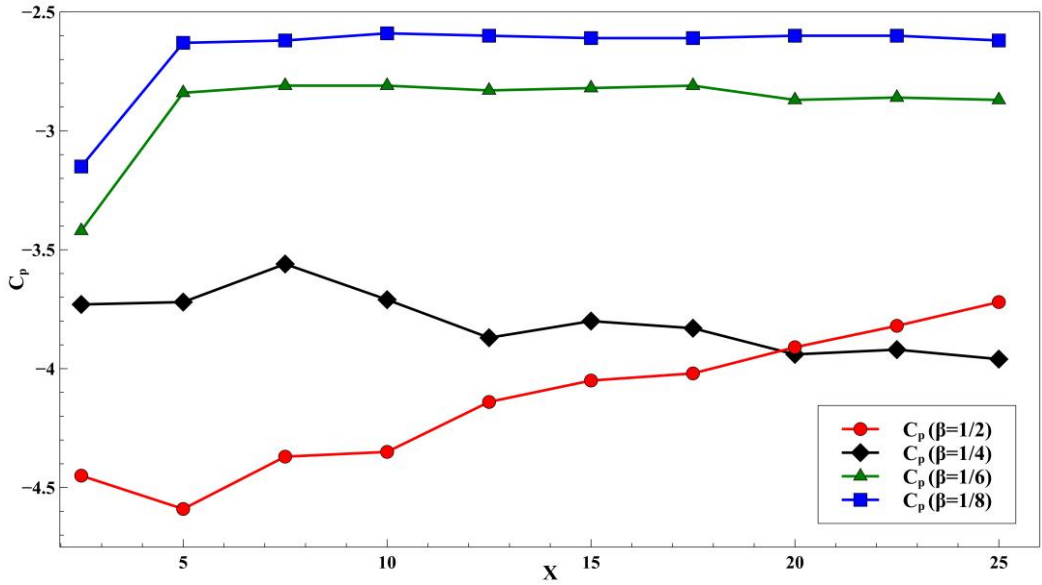

Fig 10: Variation of Pressure distribution with cross-section downstream.

In the next part we discuss the evolution and shedding of vortices in detail at higher rotational speed $\alpha = 2$.

At this value of $\alpha$, the tip of the cylinder moves at a speed that is twice as fast as that of the surrounding fluid. Figure 11 shows the time evolution of the near field vorticity contours for $Re=50$, and $\beta = 1/8$, as the cylinder rotates in a CCW manner. We notice that at AOA=90° (Fig. 11a), the cylinder is nearly surrounded by positive vorticity, with a long tongue of positive vorticity protruding towards the leeward side of the cylinder. Here, leeward side refers to that side of the cylinder that is facing the downstream. Meanwhile, the advancing edge caters to a CW vortex, which we call herein as Hovering Vortex (HV). It is so called because as the cylinder continues to rotate, this HV hovers or maintains its place above the cylinder without dissipating. This happens because the HV feeds off the same sign vorticity carried by the incoming advancing edge (Fig. 11d).

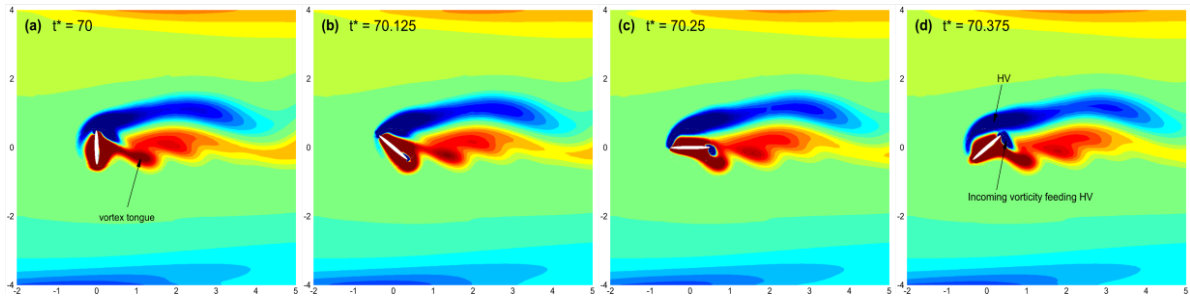

Fig 11: Instantaneous Vorticity contours at $Re = 50$, $\beta = 1/8$, $\alpha = 2$. The time instants denoted in the figures are non-dimensionalized with the rotating period of the cylinder.

A small CCW vortex shed from the retreating edge during every half cycle of the cylinder rotation. However, it takes 2.75 rotation to shed the HV from advancing edge. Also, the CW vortex shed from the advancing edge stretches out and dissipates very rapidly.

Similarly, at $Re=150$ and $\beta =1/8$, the cylinder is surrounded by a significant amount of positive vorticity. A vorticity tongue with a positive sign protrudes out near the retreating edge. Every half cycle of rotation, a small CCW vortex disengages from this tongue, and gets deflected towards the upper wall of the channel due to cylinder's rotational inertia. And the HV at advancing edge sheds after 2.25 rotation periods. Ruifeng et al [12] posited an explanation for the development of HV by parameterizing the ratio of the 'vortex shedding time scale ($T_s$)' to the 'vortex convection time scale ($T_c$)'. They defined this ratio as η. In line with Ruifeng's argument, for this particular case, we approximated the shedding time scale ~ 0.5 rotation period, and the convection time scale, measured by tracking the eye of the detached vortex as it moves in the downstream, as 2.125 rotation periods. This results in η ≈ 0.24, which is less than 1. It serves as an explanation for the merger of several (in this case, 4 vortices) CCW vortices into a single large vortex before its subsequent convection.

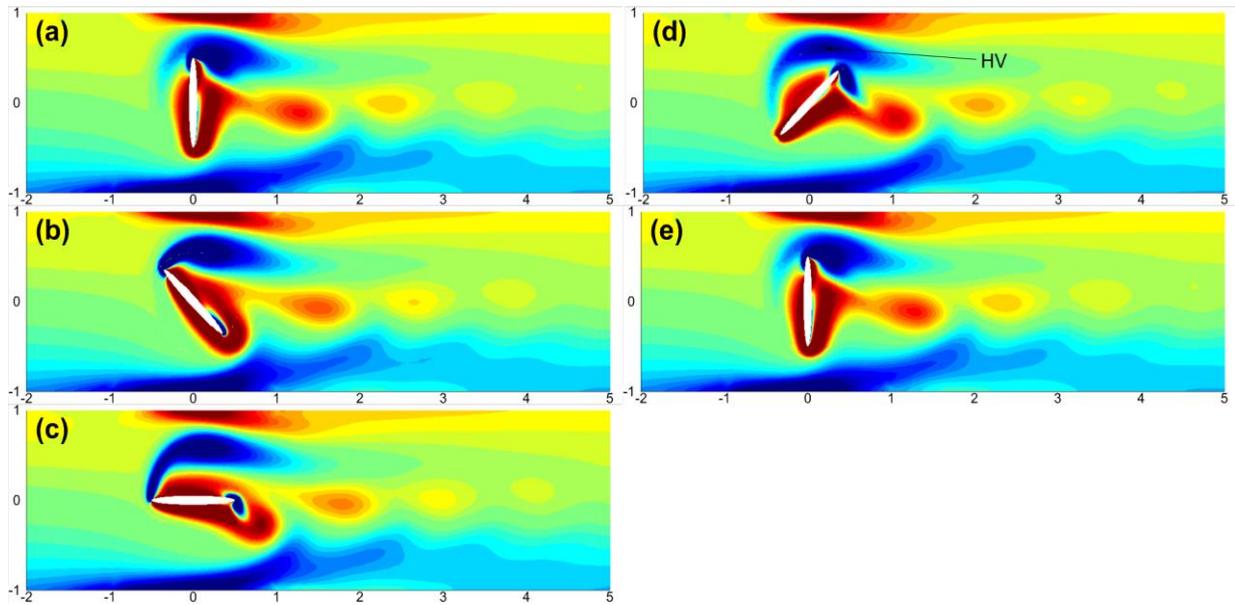

Fig 12: Vorticity Contours of $Re = 50, \beta = 1/2, \alpha = 2$ for a single vortex shedding period.

At $\beta = 1/2$, Re=50, and $\alpha = 2$, no vortex sheds from the advancing edge of the cylinder, whereas, small CCW vortices shed from the retreating edge after every half rotation period. These vortices do not make it past the $x = 5a$ mark, i.e., they are dissipated very quickly. A hovering vortex is formed above the cylinder (Fig 12(d)), which maintains its position and intensity without dissipating.

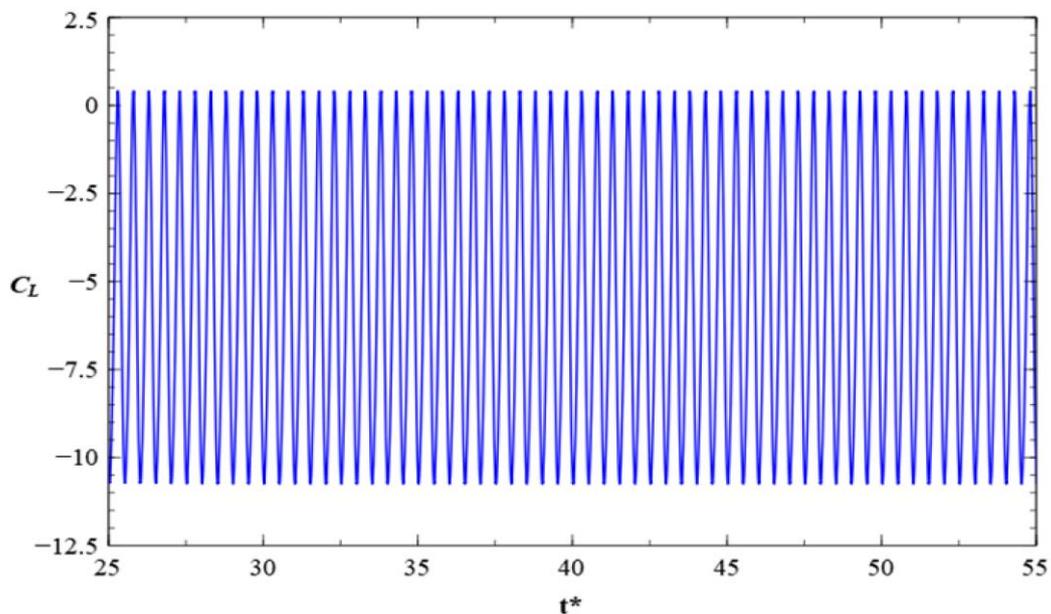

Fig. 13: Periodic lift force signal for $Re = 50$, $\beta$= 1/2, α=2.

Figure 13 shows the time variation of lift, wherein we notice that the signal is periodic after every half cycle. In addition, there is no change in the amplitude or wavelength of the signal over a span of 30 rotations (or 60 half cycles), as depicted in Figure 13. Thus, we can safely say that the Hovering vortex formed above the cylinder is 'caught' in between the wall and the cylinder, with the overall flow being quasi-steady. HVs have been reported [8,9,10,11,12] previously, but they usually shed after a finite number of cycles. In our case, the HV does not shed at all, and remains spatially stagnant. To the best of authors knowledge, this kind of behavior has not been reported before, and can be attributed to the presence of 'confinement'. We theorize that the amount of HV vorticity diffused in a half cycle is exactly compensated by the incoming vorticity of the advancing edge.

Figure 14 shows the vortex-shedding for $Re = 150$, $\alpha = 2$ and, $\beta = 1/2$. As seen before, the shedding occurs after every half rotation however, it takes two complete rotations of the cylinder to repeat the flow phenomena around the cylinder.

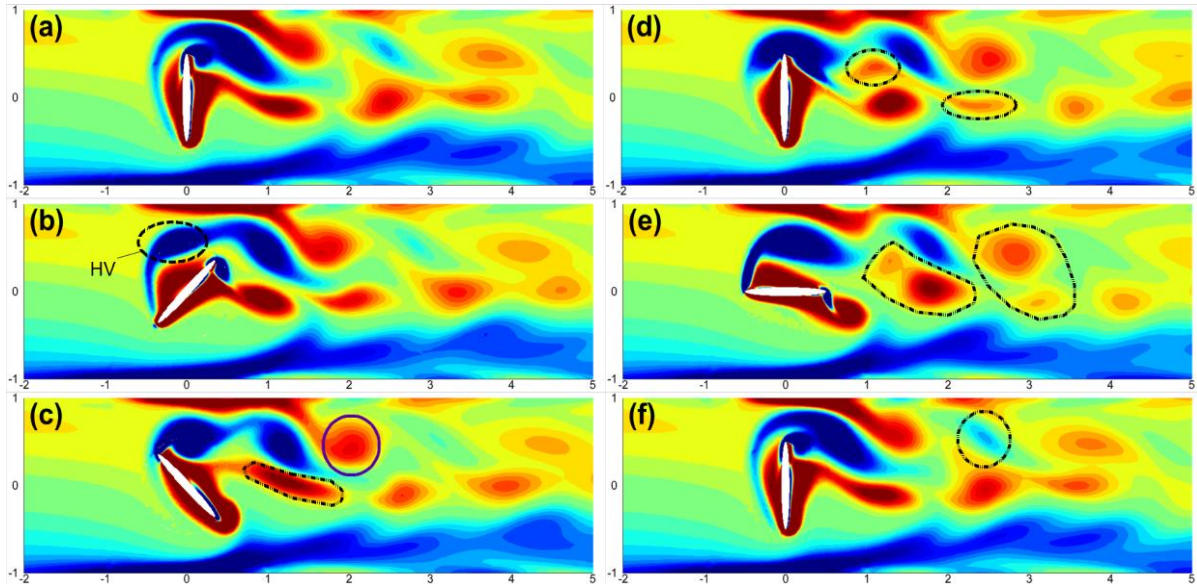

Fig 14: Instantaneous vorticity contours for $Re = 150$, $\beta = 1/2$ at instances (a)T*=124.25; (b) T*=124.625; (c) T*=124.875; (d) T*=125.25; (e) T*=125.5; (f) T*=126.25.

Because of such high rotational rate, and intense blockage, vortices disappear quickly down the stream. Unlike $Re=50$, the hovering vortex formed here (Fig 14(b)) diffuses after two complete rotations or 4-half cycles. Meanwhile, CCW vortices are shed from the cylinder after each ½ cycle. Another development that takes place is the formation of a wall vortex, shown by a solid black line in figure 14(c). From figure 14(c), we notice that the CCW vortex that is shed from the cylinder's leeward side, starts to stretch out. Finally, we observe that figure 14(f) is a clone of figure 14(a), thereby marking the end of one periodic flow sequence.

### D. Force Signals

In this section the variation of force signals with time and their FFTs have been analyzed in an attempt to explain the flow physics in a quantitative manner. Fig. 15 shows variation of instantaneous values of drag and lift coefficients with developed flow time for six complete rotations. Instances (i), (ii), (iii), (iv) and (v) are marked to relate the lift coefficient values to vorticity contours in Fig. 5-8. All the cases demonstrate a periodic variation of force signals with time. For $\beta = 1/8$, at $Re = 50$ lift force signals or flow phenomena is repeated after every half rotation period. However, the repetition time at $Re = 150$ is one full rotation of the cylinder. As mentioned before, at $Re = 50$ the same pair of vortices shed in both half period of cylinder rotation whereas the pair is slightly different (in shape and size) at $Re = 150$.

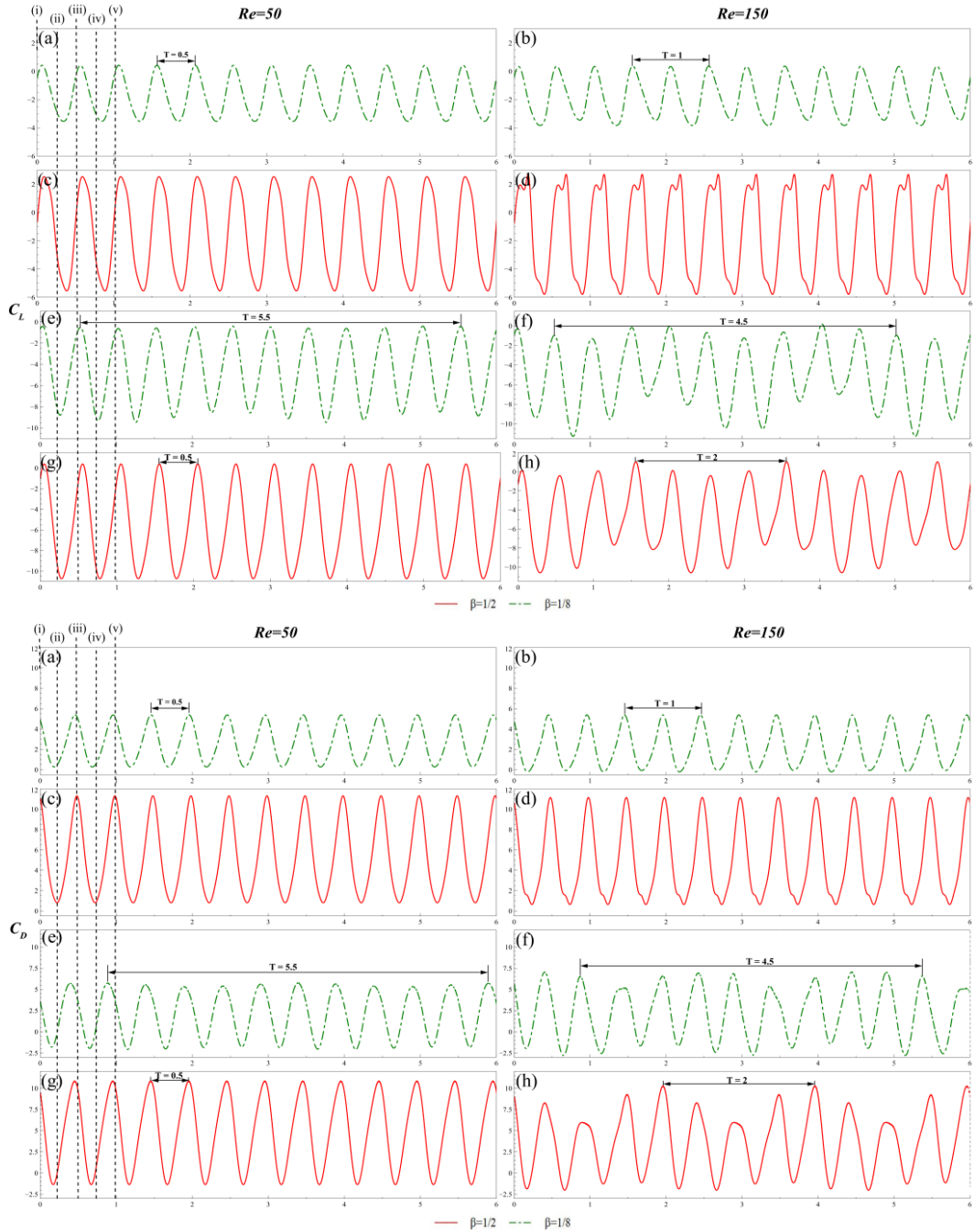

Fig 15(i) and 15(ii): Instantaneous drag and lift coefficients at $Re = 50$ & $Re = 150$ respectively. Cases (a), (b), (c) and (d) demonstrate rotation speed $\alpha = 0.5$; Cases (e), (f), (g) and (h) demonstrate rotation speed $\alpha = 2$.

With increase in blockage ratio, the flow around the cylinder is repeated after every half rotation. The vortex shedding frequency and rotation frequency remain same at $\alpha = 0.5$ for all $\beta$ values considered here. The increase in rotation speed brings significant change in time period/ frequency of flow phenomena due to presence of HV. For $\beta = 1/8$ the lift force signals repeat after 2.75 and 2.25 cylinder rotations (Fig 11) at $Re = 50$ and 150 respectively. The HV makes the shedding from advancing edge slower which leads to increase in time period after which flow phenomena gets repeated. At $\beta = 1/2$ and $Re = 50$ as discussed before, the vortex from advancing edge never sheds. The vortices at retreating edge detach and flow in downstream direction every half rotation period. Since HV never sheds off at the advancing edge, the flow phenomena in the vicinity of cylinder and lift force signals repeat after half rotation period. However, at $Re = 150$, the HV sheds after two rotations of the cylinder. In order to get insight of different frequencies in flow phenomena FFT of lift signals are observed and presented in Fig 16. The dominating peak at $f^* = 2$ in all frames signifies the rotating frequency and vortex shedding frequency. Since the orientation of cylinder is repeated after half-rotation, the rotating frequency is 2. And since a vortex always shed from retreating edge in half-rotation period, the shedding frequency is 2. Other peaks in FFT may correspond to frequencies related to flow phenomena around the cylinder and shedding from channel walls. The amplitude of force signals do not change significantly with change in Reynolds number and blockage whereas a significant variation can be seen with an increase in $\alpha$.

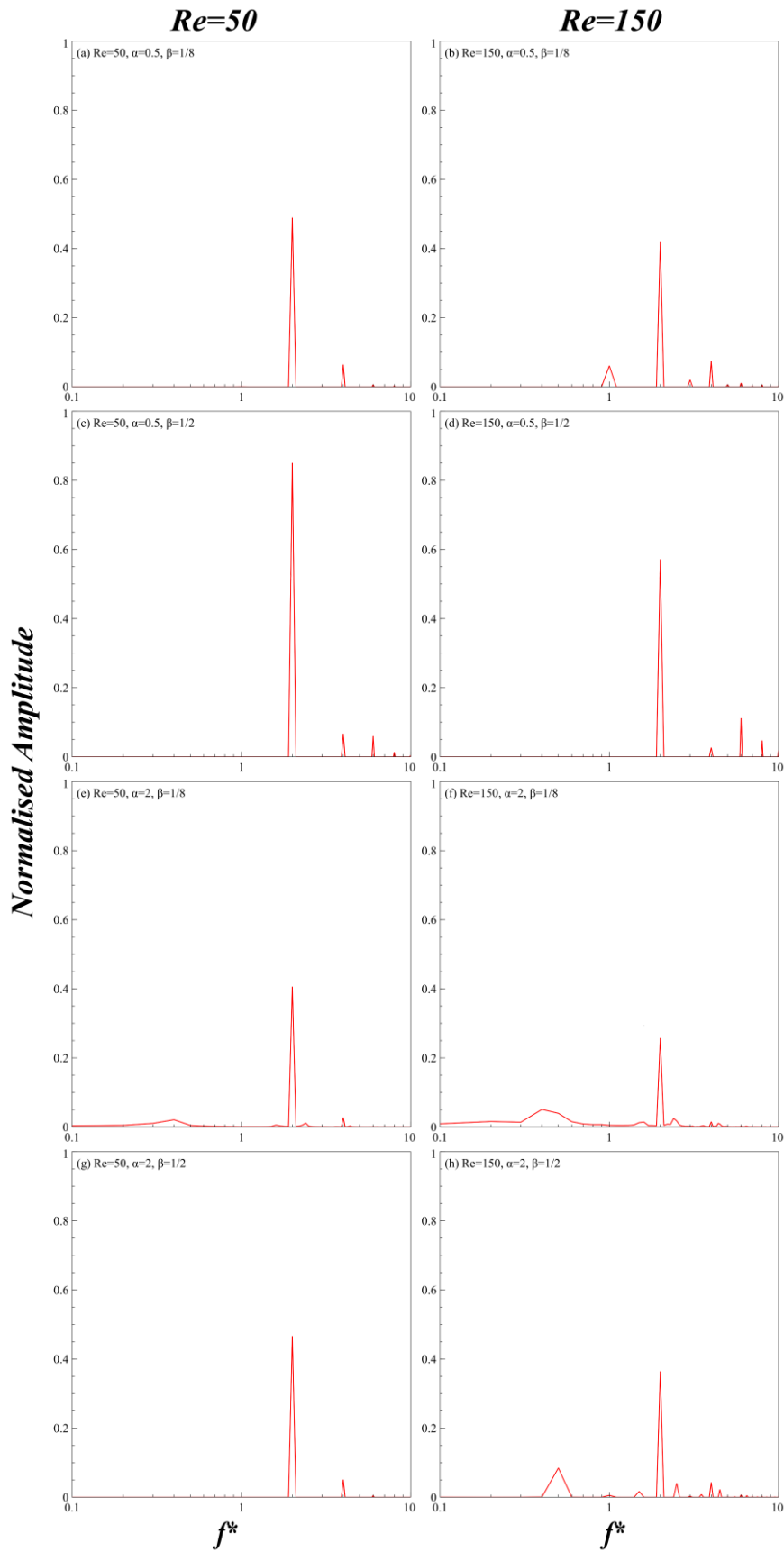

Fig 16: Fast Fourier Transforms.

E. Force Coefficients

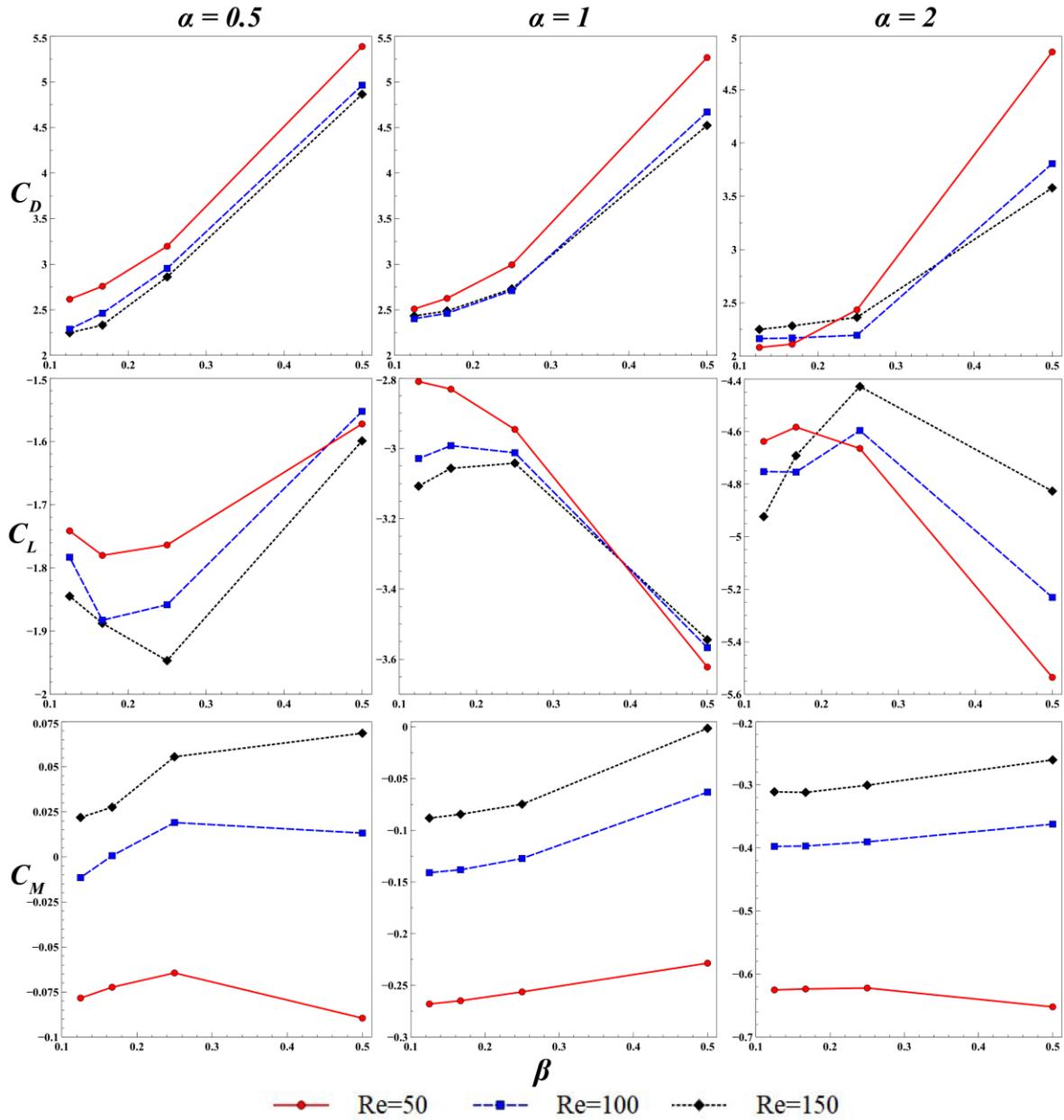

Fig 17: Variation of Time-averaged force coefficients with blockage.

The variation of time-averaged force coefficients with blockage and Reynolds number, for different values of non-dimensional rotation rate, has been shown in Fig. 17. Drag coefficient shows a monotonic increase with increase in blockage for all cases. A sharp rise in value of drag is observed as the blockage increases from 1/4 → 1/2. This rise can be attributed to interaction of rotating cylinder with viscous wall shear layers due to increase in confinement. Moreover, the value of drag coefficient decreases with increase in Reynolds number and non-dimensional rotation rate due to decreasing viscous forces, which is consistent with previous studies [18].

The variation in lift coefficient is non-monotonic at all rotation rates. This phenomenon can be explained with the help of attractive and repulsive components of lift force. Conventionally, "attractive" or negative lift occurs due to presence of shear (Z-Vorticity) near the body whereas "repulsive" or positive lift is caused by deflection of wake away from body due to the effect of confinement. The "repulsive" component is further enhanced by interaction of shed wake vortices with wall shear layers [13]. For $\alpha = 0.5, \beta = 1/2$, the deflection of wake vortices and interaction of wake vortices and wall shear layers (Fig. 7 and Fig. 8) leads to a significant repulsive component of lift force. This results in a steep rise of lift coefficient, as blockage increases from 1/4

→ 1/2 for all *Re* at $\alpha = 0.5$. At the same rotational speed, no interaction between wall shear layer and wake vortices occurs at $\beta = 1/8$ which leads to a lower value of lift coefficient. For $\alpha = 2, \beta = 1/2$, a hovering vortex is formed (Fig.12 and Fig. 14) due to increased rotation rate. Presence of this shear near the rotating cylinder leads to a steep descent of lift coefficient, as blockage increases from 1/4 → 1/2, due to dominating attractive component of lift force. It is hypothesized that there exists a specific blockage ratio at each rotation rate and Reynolds number, at which there's a switch of dominant components and one component of lift force prevails over the other, leading to either a convex or concave curve. This hypothesis can be used to explain the "U" shaped trend at rotation rate $\alpha = 0.5$ and an inverse "U" shape trend at $\alpha = 2$.

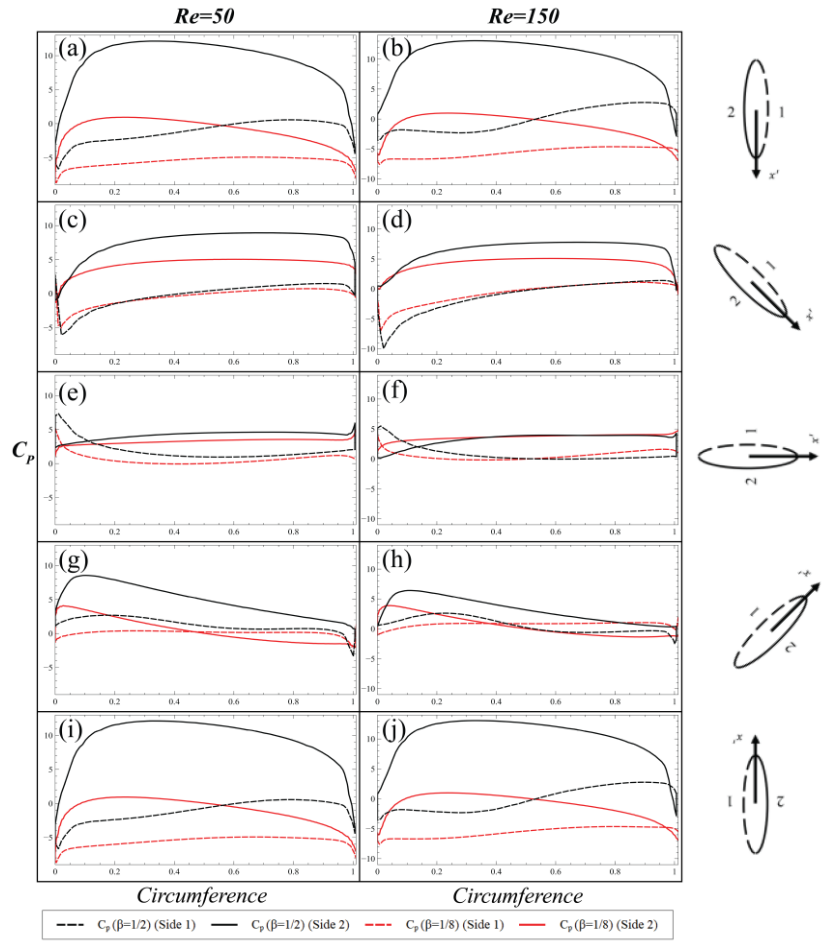

Fig 18: Variation of pressure coefficients over cylinder surface.

Figure 18 shows the variation of surface pressure coefficients over the cylinder for $\alpha = 0.5$ and $\beta = 1/2$ and 1/8 for half-rotation cycle. The higher magnitude of pressure coefficient on the Side 2 of the cylinder demonstrates the effect of incoming flow. At an orientation of $\frac{\pi}{2}$ rad, both sides show a similar range.

Time-averaged moment coefficients show a monotonic increase with increase in Reynolds number for all rotation rates. Comparing the values of moment coefficient across different rotation rates, a decreasing trend can be observed for each Reynolds number. The following section discusses the autorotation phenomenon with the help of moment coefficient.

F. Autorotation

As discussed in the introduction, 'autorotation' is a condition when the elliptic cylinder (with a fixed axis in this case) will rotate without any external torque, i.e., it will harness the energy of the flow to spin around its axis of rotation. Figure 19(a) shows the 'Riabouchinsky Curve' [2] for autorotation which depicts a non-linear relationship between the average torque ($\bar{T}$) on the Y-axis, and the angular velocity ($\Omega$)/ non-dimensional rotation rate ($\alpha$) on the X-axis. The shaded region in this $\bar{T}$ vs $\alpha$ curve represents the autorotative region. It was initially devised by Dimitri Riabouchinsky [34] based upon his autorotation experiments with a Lancaster propeller. Fixing the geometric properties of the propeller, he observed that for certain values of $\Omega$, the outside torque had a braking effect on the rotation of the propeller, i.e., it halted the rotation. He called this range 'autorotative' (shown by the shaded region). Whereas, if the values of $\Omega$ were too low or too high, the object needed external/outside torque to sustain its rotation. In this way, he performed several experiments and obtained a trend for $\bar{T}$ vs $\Omega$. It was mentioned in force coefficients section that the time-averaged moment coefficient ($C_M$) values become positive at $\alpha = 0.5$, most noticeably for $Re = 100$ and 150. As discussed by Hu and Tang [12], a positive value of $C_M$ (as per their sign conventions) was indicative of autorotation. We want to highlight that we have followed Hu and Tang's sign convention in our study, and accordingly computed the moment coefficients. Table 5 displays all the cases where $C_M$ is positive. A necessary (but not sufficient) condition for autorotation is the synchronization of vortex shedding rate and the rotational rate of the cylinder. This condition is easily achieved at $\alpha = 0.5$, and $Re = 100$ and 150. Fig. 16 showcases the Fast Fourier Transform (FFT) of the lift coefficient signal for $\beta = 1/2$ and $\beta = 1/8$. At both the values of blockage, the shedding frequency (S.F.) and rotational frequency (R.F.) coincide, which is why $C_M$ is positive for both these cases.

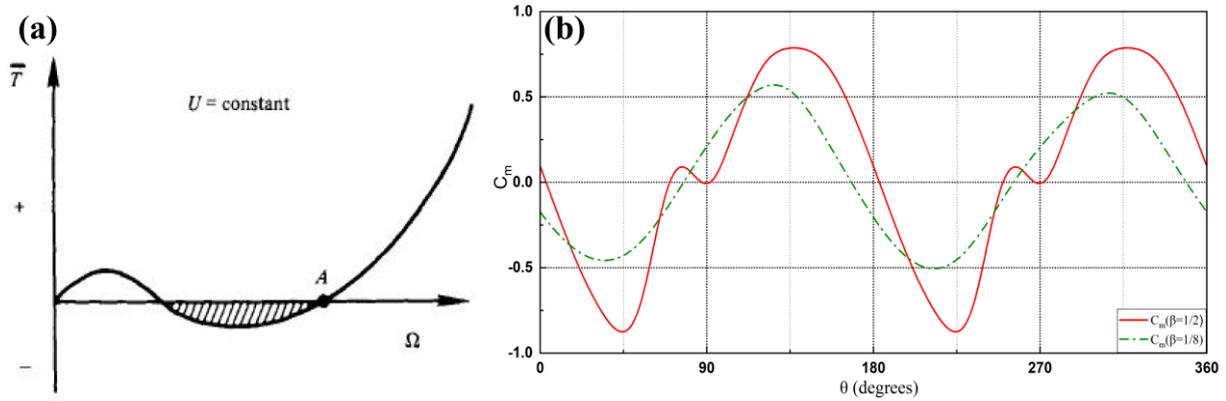

Figure 19: (a) The Riabouchinsky curve for autorotation phenomena (taken from [32]) (b) $C_M$ versus AOA curve at $Re = 150$ showing variation of moment coefficient with rotation of cylinder.

Prior studies have attempted to identify the range of $\alpha$ values for which the outside torque on the cylinder will have a braking effect and have tried to recreate the 'Riabouchinsky curve' in doing so. Based on his computations, Lugt [32] identified this auto-rotative range to start from $\alpha = 0.167$ and end at $\alpha = 0.45$ for $Re = 200$. Hu and Tang [12] recreated an inverse Riabouchinsky curve and showed that the autorotative range expands with increase in $Re$ ([12], Fig 12(c)). In their study, they did not observe any autorotation for $Re = 100$ for the considered range of $\alpha$ (0.1 – 1). However, table 5 displays the apparent benefits of introducing blockage, i.e., we are able to achieve positive values of $C_M$ at $Re = 100$ by increasing the confinement. It can thus be argued that confinement promotes autorotation, given that $Re$ and $\alpha$ lie in the feasible region.

Table V: Autorotation was observed for $\alpha= 0.5$ for the displayed $Re$ and $\beta$

| $Re$ | $\beta$ | $C_M$ |
|---|---|---|
| 100 | 1/2 | 0.01319 |
| 100 | 1/4 | 0.01905 |
| 100 | 1/6 | 0.00064 |
| 150 | 1/2 | 0.06861 |
| 150 | 1/4 | 0.05553 |
| 150 | 1/6 | 0.02755 |
| 150 | 1/8 | 0.02188 |

Figure 19(b) presents the time-variation of $C_M$ with the angle of attack ($\theta$). The part of the curve that lies above the X-axis corresponds to the region of retarding torque ($C_{MR}$), and the part lying below the X-axis is the region of supporting torque ($C_{MS}$). Where, $C_{MR}$ and $C_{MS}$ are given as:

$$C_{MR} = \frac{1}{2\pi} \int_{2\pi n}^{2\pi(n+1)} \begin{Bmatrix} C_M \ for \ C_M > 0 \\ 0 \ for \ C_M \leq 0 \end{Bmatrix} d\alpha \qquad (7)$$

$$C_{MS} = \frac{1}{2\pi} \int_{2\pi n}^{2\pi(n+1)} \begin{Bmatrix} C_M \ for \ C_M \leq 0 \\ 0 \ for \ C_M > 0 \end{Bmatrix} d\alpha \qquad (8)$$

Detailed discussions and definitions for retarding and supporting torque can be found elsewhere [32].

## V. CONCLUSION

This paper numerically investigates the effect of rotation for an elliptic cylinder in a confined channel. The parameter space sweeps through $50 \leq Re \leq 150$, $0.5 \leq \alpha \leq 2$ and $1/8 \leq \beta \leq 1/2$. The aspect ratio of the cylinder is kept constant throughout the study. Both qualitative (in form of vortex shedding) and quantitative (in form of force coefficients) analysis of results illustrates dependence of flow physics on all the parameters. The vortex shedding phenomena has been extensively discussed, in both near and far-fields to discuss the effect of flow inertia, confinement and rotation. Weakening of vortex shedding is observed with increasing $\alpha$ and $Re$ and with decreasing $\beta$. While the drag force coefficients show a monotonic increase with blockage, the non-monotonic variation of lift force coefficients demonstrates the effect of gap-flow and interplay of attractive and repulsive components across the blockage parameter space. Further studies can be performed to examine complete suppression of vortex shedding. Autorotation phenomena was favored at lower values of $\alpha$, and at higher values of $\beta$ and $Re$. This phenomenon can be further investigated for increased confinement and flow inertia. In addition, compressible flows can be considered in future works.

## VI. Acknowledgement

The authors would like to thank Ms. Chinu Routa for several discussions and valuable insights that have led to this paper. The authors would further like to acknowledge the essential computational resources provided by the Department of Chemical engineering, National Institute of Technology Rourkela.

List of table captions

List of figure captions

- Fig 1: Schematic representation of computational domain.
- Fig 2: Schematic representation of Multi-block mesh used in the study.
- Fig 3: Streamlines of $Re = 150, \beta = 1/2, \alpha = 0.5$.
- Fig 4: Streamlines of $Re = 150, \beta = 1/8, \alpha = 0.5$.
- Fig 5: Vorticity Contours of $Re=50, \beta=1/8, \alpha=0.5$ for a single vortex shedding period.
- Fig 6: Vorticity Contours of $Re=150, \beta=1/8, \alpha=0.5$ for a single vortex shedding period.
- Fig 7: Vorticity Contours of $Re=50, \beta=1/2, \alpha=0.5$ for a single vortex shedding period.
- Fig 8: Vorticity Contours of $Re=150, \beta=1/2, \alpha=0.5$ for a single vortex shedding period.
- Fig 9: Far-field vorticity contours for all blockages at $Re=150$ and $\alpha=0.5$.
- Fig 10: Variation of Pressure distribution with cross-section downstream.
- Fig 11: Instantaneous Vorticity contours at $Re=50, \beta=1/8, \alpha=2$. The time instants denoted in the figures are non-dimensionalized with the rotating period of the cylinder.
- Fig 12: Vorticity Contours of $Re=50, \beta=1/2, \alpha=2$ for a single vortex shedding period.
- Fig 13: Periodic lift force signal for $Re=50, \beta=1/2, \alpha=2$.
- Fig 14: Instantaneous vorticity contours for $Re=150, \beta=1/2$ at instances (a)$T^*$=124.25; (b) $T^*$=124.625; (c) $T^*$=124.875; (d) $T^*$=125.25; (e) $T^*$=125.5; (f) $T^*$=126.25.
- Fig 15(i) and 15(ii): Instantaneous drag and lift coefficients at $Re=50$ & $Re=150$ respectively. Cases (a), (b), (c) and (d) demonstrate rotation speed $\alpha$=0.5; Cases (e), (f), (g) and (h) demonstrate rotation speed $\alpha$=2.
- Fig 16: Fast Fourier Transforms.
- Fig 17: Variation of Time-averaged force coefficients with blockage.
- Fig 18: Variation of pressure coefficients over cylinder surface.
- Fig 19: (a) The Riabouchinsky curve for autorotation phenomena (taken from [32]) (b) $C_M$ versus AOA curve at $Re=150$ showing variation of moment coefficient with rotation of cylinder.